\documentclass{svproc}
\usepackage{graphicx}
\usepackage{url}

\begin{document}
\mainmatter              % start of a contribution
\title{Gravitational Waves as a Probe of Particle Dark Matter}
%
%\titlerunning{Hamiltonian Mechanics}  % abbreviated title (for running head)
\author{Sulagna Bhattacharya\inst{1}}
%
%\authorrunning{Ivar Ekeland et al.} % abbreviated author list (for running head)
%Craig Chambers, Kim B. Bruce, and Elisa Bertino}
%
\institute{Tata Institute of Fundamental Research, Homi Bhabha Road, Mumbai 400005, India\\
\email{sulagna@theory.tifr.res.in}}

\maketitle              % typeset the title of the contribution

\begin{abstract}
Galactic dark matter (DM) particles, having non-gravitational interactions with nucleons, can interact with stellar constituents and eventually become captured within stars. Over the lifetime of the celestial body, these non-annihilating, heavy DM particles may accumulate and eventually form a comparable stellar mass black hole (BH), referred to as a Transmuted Black Hole (TBH). We investigate how current gravitational wave (GW) experiments could detect such particle DM through the presence of low-mass TBHs, which cannot form via standard stellar evolution. Different stellar objects (compact and non-compact) provide laboratories across  DM-nucleon interaction regimes, offering insight into DM's mass and its non-gravitational properties.
% We would like to encourage you to list your keywords within
% the abstract section using the \keywords{...} command.
\end{abstract}
\section{Introduction}
Cosmological and astrophysical observations over recent decades strongly support the existence of dark matter DM \cite{Planck:2018vyg}. Given its electromagnetic inertness, efforts to reveal DM's nature focus on gravitational effects and possible non-gravitational interactions with standard model particles \cite{Cirelli:2024ssz}, aiming to constrain its fundamental properties and role in the universe. Viable DM candidates span an extensive mass range, from ultralight scalars \cite{Hui:2016ltb} around $\sim 10^{-22}$\,eV to primordial black holes (PBHs) up to masses $10^{(3-5)} M_{\odot}$ \cite{Carr:1974nx,Chapline:1975ojl}, covering more than 80 orders of magnitude. In this study, we focus on non-annihilating, heavy DM particles \cite{Petraki:2013wwa,Zurek:2013wia} that, upon capture and accumulation within stars, may form low-mass black holes—a scenario incompatible with standard stellar evolution theory.

Recent gravitational wave events have raised the possibility that low-mass BHs may exist within detected binary mergers \cite{LIGOScientific:2020aai,LIGOScientific:2020zkf,LIGOScientific:2021qlt,Prunier:2023uoo}. These events include objects whose masses could represent either the heaviest neutron stars or the lightest black holes, observed so far. While PBHs are considered a potential explanation for these unconventional BHs, we propose that DM capture-induced transmutation could also produce such exotic BHs \cite{Dasgupta:2020mqg,Bhattacharya:2023stq,Ray:2023auh,Bhattacharya:2024pmp}. The following sections outline the formation mechanism of these transmuted BHs and their astrophysical implications.

\section{GW Observations Through TBHs}
\paragraph{Formation of TBHs:} Non-annihilating DM particles \cite{Petraki:2013wwa,Zurek:2013wia}, possessing a non-zero but sufficient interaction cross-section ($\sigma_{\chi n}$) with nucleons, can lose energy through single \cite{Bell:2020lmm} or multiple scattering \cite{Bramante:2017xlb} events, eventually becoming gravitationally captured within celestial bodies. Through repeated interactions with stellar constituents, these particles thermalize, forming a core of radius $r_{\rm th}$. Heavier DM particles yield a more compact core with higher density. This dark core may then collapse into a small BH due to self-gravitation \& Chandrasekhar collapse, depending on whether the DM particles are Bosons or Fermions. This nascent BH, once formed, can accrete surrounding stellar matter, and if accretion dominates over Hawking evaporation, the entire stellar object may transform into a BH of comparable mass, termed as a TBH \cite{Dasgupta:2020mqg,Bhattacharya:2023stq,Ray:2023auh,Bhattacharya:2024pmp}. The two main timescales in this scenario are $\tau_{\rm collapse}$, the time required to capture enough DM particles for dark core collapse, and $\tau_{\rm swallow}$, the time to consume the entire star, resulting in formation of the TBH. For suitable DM mass ($m_{\chi}$) and cross-section ($\sigma_{\chi n}$), if $\tau_{\rm collapse} + \tau_{\rm swallow} < t_0$ (age of the universe = $13.8 \times 10^9$ yrs), we should start detecting these BHs observationally. Non-observation imposes stringent constraints on the DM parameter space.

Neutron stars are optimal probes for weak interaction cross-sections \cite{Bhattacharya:2023stq} due to their high gravitational potential, which enhances the capture and retention of DM particles. In contrast, Sun-like (non-compact) stars are better suited to probing regimes with comparatively stronger DM-nucleon interactions \cite{Bhattacharya:2024pmp}, where capture rates are still comparable despite the lower gravitational binding energy. Table [\ref{Table}] gives some timescales involved in this scenario for specific DM mass and interaction cross-section. For NS we take $M_{\rm NS}=1.35 M_{\odot}, R_{\rm NS}=10$\,km, $T_{\rm NS}=2.1\times10^6$\,K, $T_{\rm age}|_{\rm NS}=10^9$\,yrs and for Sun-like stars $M_{\odot}=1.99\times10^{30}$\,kg, $R_{\odot}=6.9\times10^5$\,km, $T_{\rm core}=1.5\times10^{7}$\,K, $T_{\rm age}|_{\rm Sun}=4.6\times10^9$\,yrs.
\begin{table}[h!]
	\centering
	\caption{Key quantities for the TBH formation scenario. See Ref.\cite{Bhattacharya:2023stq,Bhattacharya:2024pmp} for detailed calculations.}
	\begin{tabular}{|c|c|c|c|c|}
		\hline
		&\parbox[c]{2cm}{\vspace{0.15cm}\centering{ Capture Rate \\ (no. of particles/ sec)\vspace{0.15cm}}}& \parbox[c]{4cm} {\vspace{0.15 cm}\centering{Thermalization Radius}\\ $r_{\rm th}$ in ($R_{\rm NS}/ R_{\odot}$)\vspace{0.15 cm}}&\parbox[c]{1.5 cm} {\centering{$\tau_{\rm collapse}$\\  (yrs)}}& \parbox[c]{1.5 cm} {\centering{$\tau_{\rm swallow}$\\  (yrs)}}\\
		\hline 
		\parbox[c]{3cm}{\vspace{0.1 cm}\centering{\textbf{Neutron Stars}}\\ ($m_{\chi}=10^5$ GeV, $\sigma_{\chi n}=10^{-45}\,\rm cm^2$)\vspace{0.1 cm}} & $2\times 10^{20}$ & $5\times10^{-6}\, R_{\rm NS}$& $4.8\times10^8$ & $3\times10^4$ \\
		\hline 
		\parbox[c]{3cm}{\vspace{0.1 cm} \centering{\textbf{Sun-like stars}}\\ ($m_{\chi}=10^7$ GeV, $\sigma_{\chi n}=10^{-30}\,\rm cm^2$)\vspace{0.1 cm}} & $1\times10^{23}$ & $4\times10^{-5}\,R_{\odot}$& $2\times10^8\,$& $10^4$\,yrs\\
		\hline
	\end{tabular}
	\label{Table}
\end{table}
\begin{figure}[h!]
	\centering
	\includegraphics[width=0.43\textwidth]{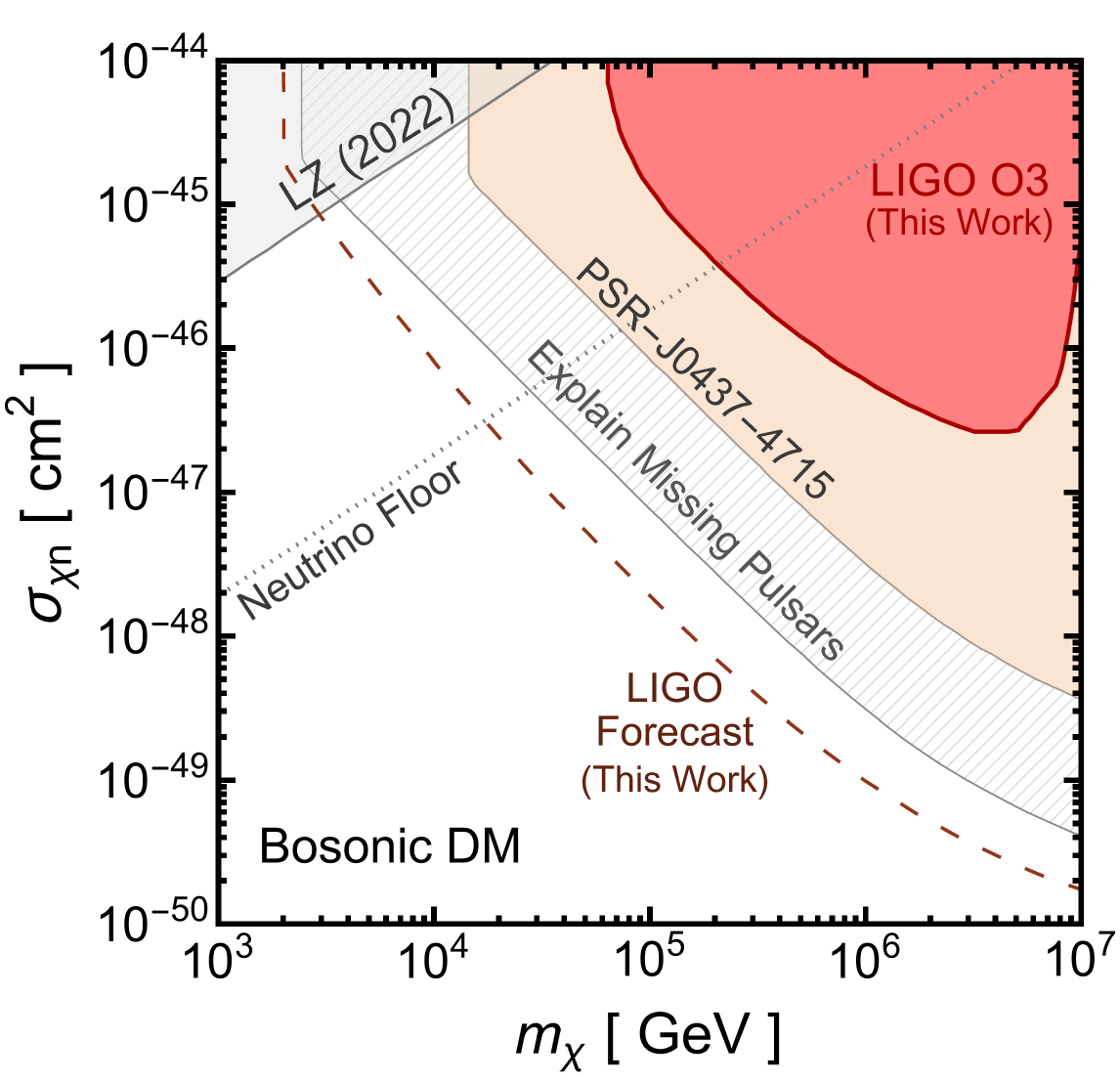}
	\hspace{1.2 cm}
	\includegraphics[width=0.43\textwidth]{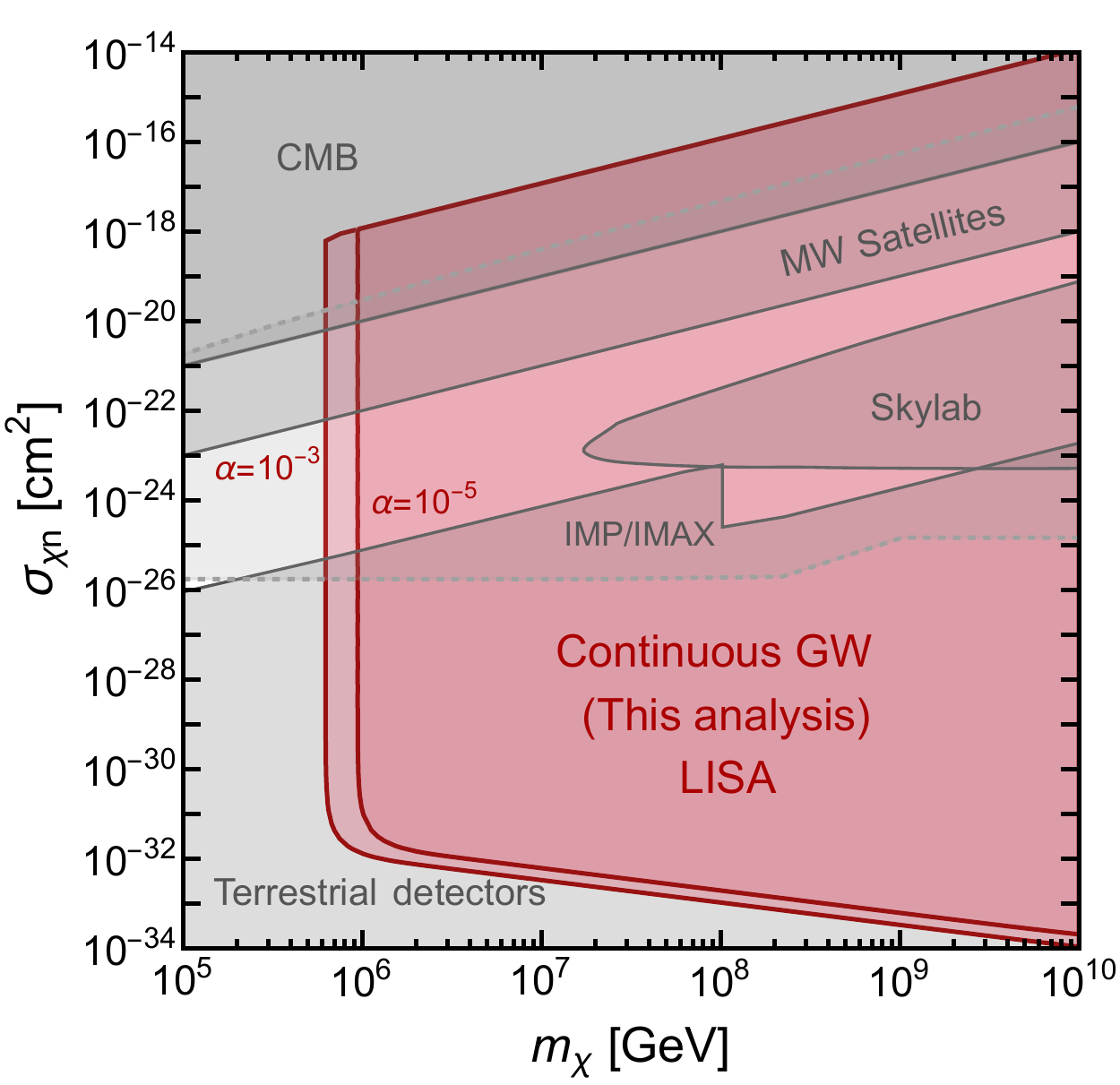}
	\caption{Exclusion in the DM mass ($m_{\chi}$) and interaction cross-section ($\sigma_{\chi n}$) plane for Bosonic DM particles. Left Panel: The red shaded region shows the current limit from LVK’s O3 observation \cite{LIGOScientific:2022hai}, while the brown dashed line represents the forecast with a 50$\times$ sensitivity increase. (See Fig.1 of Ref.\cite{Bhattacharya:2023stq}). Right Panel: The red shaded region indicates the exclusion based on LISA’s proposed sensitivity, with $\alpha$ being the close binary fraction. For $\alpha < 10^{-5}$, this method excludes no parameter space. (Figure from Ref.\cite{Bhattacharya:2024pmp}). Further details on other existing constraints and other scenarios (Fermionic DM, Bose Einstein Condensate formation) are in the original Refs.\cite{Bhattacharya:2023stq,Bhattacharya:2024pmp}.}
	\label{fig}
\end{figure}
\paragraph{Constraints from Gravitational Wave Observations:}
DM capture-induced transmutation can produce BHs with masses comparable to neutron stars, making their binary mergers detectable by the LIGO-Virgo-KAGRA (LVK) collaboration. The absence of low-mass BH observations in the LVK O3 data \cite{LIGOScientific:2022hai} imposes an upper limit on the merger rate of such events. Ref.\cite{Bhattacharya:2023stq} compares this upper limit with the theoretical TBH merger rates \cite{Dasgupta:2020mqg}, excluding parts of the DM parameter space, and sets forecast limits for further exclusions with continued non-observations in future LVK data (left panel of Fig.\ref{fig}).

Continuous gravitational waves (CGWs) are quasi-monochromatic signals with minimal frequency evolution. The Laser Interferometer Space Antenna (LISA) or Big Bang Observatory (BBO) are capable of detecting these CGWs at frequencies starting from $10^{-5}$\, Hz. Stellar binary inspirals can serve as potential sources of CGWs; however, they must be extremely close (4$R_{\odot}$-10$R_{\odot}$) binaries to fall within LISA's sensitivity band. DM particles can become trapped within stars in close stellar binaries, potentially forming similar-mass black holes (BHs) that result in close stellar-mass BH binaries. Ref.\cite{Bhattacharya:2024pmp} shows that non-observation of CGWs from such systems in LISA/BBO data can place an upper limit on their occurrence rate density, translating into exclusions in the DM parameter space, particularly in the stronger interaction regime (right panel of Fig.\ref{fig}).
\paragraph{Summary:} Existing terrestrial detectors \cite{LZ:2022lsv} lack the sensitivity to probe heavy DM particles, whereas current GW detectors offer a more effective approach. Non-observation of TBHs can set world-leading constraints on the DM parameter space, where a detection could reveal DM's true nature. These constraints can be further strengthened in the higher mass range by considering modified Hawking evaporation scenarios for nascent BHs formed within celestial cores \cite{Basumatary:2024uwo}. See Ref. \cite{Bramante:2023djs} to look for other promising avenues to unveil DM properties via compact stars.
\paragraph{Acknowledgement:} I thank Basudeb Dasgupta, Ranjan Laha, Andrew Miller, and Anupam Ray for collaborations on the original works \cite{Bhattacharya:2023stq,Bhattacharya:2024pmp} and for ongoing scientific discussions.


\end{document}